\documentclass[twocolumn,pra]{revtex4}

\usepackage{amsmath}
\usepackage{amssymb}
\usepackage{graphicx}

\providecommand{\betrag}[1]{\lvert#1\rvert}
\DeclareMathOperator{\var}{var}

\begin{document}

\title{Theory of the Photocount Statistics for Multi-Mode Multi-Frequency
Radiation Fields}

\author{Michael Patra}
\affiliation{Carl Zeiss SMT AG, 
Rudolf-Eber-Stra{\ss}e 2, 73447 Oberkochen, Germany}

\begin{abstract}

We derive on the level of quantum optics expressions for the uncertainty of the
photocount in a multi-mode multi-frequency setup. The result depends on the
quantum correlations of the individual modes and the frequency spectrum of the
radiation, the latter leading to a frequency beating sometimes referred to as
dynamic laser speckle. When the mode structure of the radiation field is disturbed
between source and detector, another contribution to the photocount uncertainty
referred to as static speckle appears. To predict the size of this effect, we
present a suitable definition of the etendue (or phase space volume) that links
the number of modes of a radiation field to macroscopic quantities.

\end{abstract}


\maketitle

\section{Introduction}

Given some radiation field, a prediction can be made about the outcome of an
experiment where a suitable detector is placed at some position $\vec{R}$ and
probes the radiation field for some time $T$. Even when the radiation field is
known as precisely as possible, there is still a finite uncertainty in the
outcome. This statement can also be formulated in the inverse way: it is not
possible to build an apparatus that provides a ``better'' (=more defined)
illumination at the point $\vec{R}$ than this limit.

We will demonstrate in this paper that the uncertainty of the photocount,
quantified by its variance, depends on both the number of modes, including the
energy distribution among the modes, and on the spectral properties and  quantum
correlations of the radiation field. Another contribution to the uncertainty
will appear if the radiation field is known at the source but the mode structure
is then perturbed in some uncontrolled way before it reaches the location of the
detector. The latter effect is frequently referred to as static
speckle~\cite{goodman:07} named after the pattern it creates on a camera.  The
opposing term ``dynamic speckle'' unfortunately is used to denote two completely
different concepts in the literature.
Frequently it refers to a static speckle pattern that is changing over
time because the source of the perturbation is moving~\cite{fainman:81a} but 
in this paper
we refer to
``dynamic laser speckle''~\cite{rydberg:06a} -- an effect that is 
due to the quantum dynamics of the radiation field.

Quantum effects are most prominent on small length scales. Modern optical
lithography operates precisely in this regime, printing structures smaller than
the wavelength of the light with a precision of a few
nanometres~\cite{mack:07}. To this end, photo resists have been developed that
are more sensitive to changes of the light intensity than most technical
sensors, thereby acting as efficient (albeit unintentional) detectors for
uncertainties of the photocount. Later in this paper we will show that the
effects of multi-mode multi-frequency radiation  are most pronounced neither at
the limit of large or short times but in the intermediate regime. Recent
studies have confirmed that this is precisely the regime used in modern optical
lithography~\cite{noordman:09a}.

\section{Overview}

Classically a radiation field is described by its electric field
$\vec{E}(\vec{r},t)$ as function of position $\vec{r}$ and time $t$. In a
quantum treatment, the electric field $\vec{E}$ is replaced by a
suitable operator. In both treatments, the electric field is not a perfectly
determined quantity, and the maximum knowledge possible is contained in the
mutual-coherence function~\cite{mandel:95}
\begin{equation}
	\Gamma(\vec{r}_1,\vec{r}_2,t_1,t_2)=\langle \vec{E}^*(\vec{r}_1,t_1)
		\vec{E}(\vec{r}_2,t_2) \rangle \;.
	\label{eqGamma}
\end{equation}
We restrict ourselves to stationary fields such that $\Gamma$ does not depend
on $t_1$ and $t_2$ but only on the time-difference $t_1-t_2$. Furthermore, for
ease of writing we
will only treat one component $E$ of the electric field $\vec{E}$ but the
extension to cover the other components is straight-forward.

The Fourier transform $\hat{\Gamma}(\vec{r}_1,\vec{r}_2,\omega)$ of 
$\Gamma(\vec{r}_1,\vec{r}_2,t_1-t_2)$ with respect to the time difference
is a nonnegative Hermitian
operator, and thus possesses an eigenrepresentation~\cite{mandel:95}
\begin{equation}
	\hat{\Gamma}(\vec{r}_1,\vec{r}_2,\omega)=\sum_n \alpha_n(\omega)
	\phi^*_n(\vec{r}_1,\omega) \phi_n(\vec{r}_2,\omega) \;,\qquad
	\alpha_n(\omega) \ge 0 \;.
	\label{eqGamma2}
\end{equation}
The eigenvectors $\phi_n(\vec{r},\omega)$ are called the modes of the electric 
field and form a complete orthonormal set. We assume that the frequency 
spectrum is small enough such that $\phi_n(\vec{r},\omega)$ is independent 
of $\omega$. 
Whenever the frequency spectrum is equal to the natural linewidth of the light
source, i.\,e., the spectrum is due to the finite lifetime of some excited
light-emitting medium such as in a laser, this assumption is always fulfilled.
For other light sources, this assumption can sometimes be problematic 
for cavity-like
systems but for open system or in a waveguide geometry, this is less of
an issue.
If the condition that $\phi_n(\vec{r},\omega)$ is independent 
of $\omega$, should be violated, one can split the mode into several 
discrete modes (one for each frequency interval in which
the shape of the mode can assumed to be constant) and the remainder of this 
paper be applied
nonetheless. The electric field can then be written as
\begin{equation}
	E(\vec{r},t)= \sum_n a_n(t) \phi_n(\vec{r}) \;.
	\label{eqGamma3}
\end{equation}
Since $\phi_n$ is independent of $t$, the $\phi_n$ are the modes instead 
of being some arbitrary base of the electric field. Comparison of
Eqs.~(\ref{eqGamma2}) and~(\ref{eqGamma3}) shows that the magnitude of $a_n$ can
be computed from $\alpha_n$ but not its phase. A semiclassical treatment of
the photocount statistics would assume that $\alpha_n$ is fluctuating in time
whereas within quantum optics, which we will apply in this paper,
fluctuations are inherent to the operator description.

This paper is organised as follows. In Sec.~\ref{secZeitunschaerfe} we compute
the variance of the photocount on a quantum-optical level  when the radiation
field is completely known. We will find that there is a shot noise term, a 
quantum correlation term and a term describing the beating of different
frequencies. Frequently, by design or unintentionally, the mode structure
emitted by a known light source is completely changed and unknown when it
reaches the point of the detector. As will be demonstrated in
Sec.~\ref{secStaticSpeckle}, random-matrix theory allows a compact and exact
treatment of this problem. The resulting ``static speckle'' becomes the smaller
the more modes of the radiation field are excited, and the more uniform the
energy distribution among the modes is. Depending on the ``size'' of the
radiation field, there is thus a minimum amount of static speckle, which is
computed in Sec.~\ref{secEtendue}. In Sec.~\ref{secStaticKorrelation} we extend
this question to computing the most likely amount of static speckle. Since we
will demonstrate this quantity to be self-averaging, the computed average is
more than just an average in that it is characteristic for almost all 
individual speckle values.

\section{Quantum theory of photodetection}
\label{secZeitunschaerfe}

While in ``general'' electrodynamics the electric field $\vec{E}(\vec{r},t)$
can be probed directly, this is not possible in the realm of optics as
$\vec{E}(\vec{r},t)$ changes too quickly in time and space to allow a direct
measurement. Rather, the radiation field is probed by means of photodetection,
i.\,e., by absorbing photons inside some device and counting the number of
photons absorbed. This principle applies to technical machines as well as 
to the human eye or to photo resists.

On small length and time scales, quantum effects become important. For this
reason, and because all effects can then be treated in a more compact way, we
will use the quantum theory of photodetection in the following. This
theory~\cite{glauber:63a,kelley:64a} is usually formulated for single-mode
detection. Since multi-mode fields~\cite{fleischhauer:91a} are at the core of
this paper, we will present an extension of the theory to multi-mode
photodetection here. We will treat the multi-frequency aspect explicitly and
not express the different frequencies by different modes as is frequently
done in textbooks.

We label the modes of the electromagnetic field by $\phi_n(\vec{r})$. The
annihilation operator associated with this mode is $a_n$. 
Using this notation, the quantum operator for the electric field at
some position $\vec{R}$ is given by~\cite{mandel:95}
\begin{equation}
	F(t)= \sum_n \phi_n(\vec{R}) a_n(t)\;.
	\label{eqFexpansion}
\end{equation}
Later it will prove helpful to switch from the time representation $a_n(t)$ to
the spectral representation $a_n(\omega)$,
\begin{equation}
	a_n(t)=\frac{1}{\sqrt{2\pi}}\int a_n(\omega) \mathrm{e}^{-\mathrm{i}\omega t}
			\mathrm{d}\omega\;.
	\label{eqAfrequenz}
\end{equation}
since in time representation, operators taken at different times do not
necessarily commute whereas they do in the spectral representation,
\begin{equation}
	[a_n(\omega),a_m^{\dagger}(\omega')]=\delta_{nm}\delta(\omega-\omega')
	\;.
	\label{eqVertauschungsrelation}
\end{equation}	
When we label the number of photons in the $n$-the mode with $I_n$, 
which basically amounts to the total intensity of that mode,
this gives the expectation value
\begin{equation}
	\langle a_n^\dagger(\omega) a_m(\omega) \rangle = 2 \pi \delta_{nm}
		\delta(\omega-\omega') I_n G_n(\omega)\;,
	\label{eqErwart1}
\end{equation}		
where $G_n(\omega)$ is the (normalised) spectrum of the radiation in the $n$-th
mode, and the prefactor $2\pi$ has been introduced for later convenience.

Photodetection within some time interval $T$ is then described by the 
quantity~\cite{mandel:95}
\begin{equation}
	W = \eta \int_0^T F^{\dagger}(t) F(t) \mathrm{d}t\;,
\end{equation}
where $\eta$ marks the detection efficiency and includes information on the size
of the detector. This amounts to a perturbative description of the interaction
between the detector and the radiation field. The perturbative approach neglects
the effect that every detected photon decreases the number of photons remaining
in the radiation field~\cite{fleischhauer:91a} but this is mainly an issue for
microcavities where only a small number of photons are excited at one point in
time and the dynamics are then studied. For the purpose of this paper, this is
no relevant restriction.

The factorial moment $n^{(k)}:=\langle n (n-1) \cdots
(n-k+1)\rangle$ of the photodetection count distribution is given by
\begin{equation}
	n^{(k)} = \langle : W^k : \rangle \;,
	\label{eqFactorialMoment1}
\end{equation}
where the colons denote normal-ordering of the operator within, and the brackets
$\langle \ldots \rangle$ denote the average which has to be taken over both the
quantum fluctuations of the operators $a_n$ and the frequency distribution
$G_n(\omega)$. Mean and variance of the photo count then follow from
\begin{equation}
	\overline{n}=n^{(1)}\;,\qquad
	\var n=n^{(1)}+n^{(2)}-\bigl[n^{(1)}\bigr]^2\;.
	\label{eqDefVar}
\end{equation}	

For the first moment, this yields the result
\begin{multline}
n^{(1)}=
\eta \langle \int_0^T \mathrm{d}t
	\sum_{nm} \phi_n^*(\vec{R})\phi_m(\vec{R})a^{\dagger}_n(t) a_m(t) \rangle
	= \\ 
	\frac{\eta}{2\pi}\int_0^T \mathrm{d}t
	\int\!\!\!\!\int\mathrm{d}\omega_{1,2}
	\sum_{nm} \phi_n^*(\vec{R})\phi_m(\vec{R})
	\langle a^{\dagger}_n(\omega_1) a_m(\omega_2) \rangle
	\mathrm{e}^{\mathrm{i}(\omega_1 - \omega_2) t} \\
	= \eta \int_0^T \mathrm{d}t \int\mathrm{d}\omega \sum_n
		 \betrag{\phi_n(\vec{R})}^2 I_n G_n(\omega)
	= T \eta \sum_n \betrag{\phi_n(\vec{R})}^2 I_n
	\;.
	\label{eqAfrequenz2}
\end{multline}
This is the same result as expected by classical theory since
$\betrag{\phi_n(\vec{R})}^2 I_n$ corresponds to the intensity of the $n$-th mode
at position $\vec{R}$. 

The second factorial moment is
\begin{widetext}
\begin{multline}
	n^{(2)}=\eta^2 \langle :\int\!\!\!\int_0^T\mathrm{d}t_{1,2}
		\sum_{n_1,\ldots,n_4} \phi_{n_1}^*(\vec{R}) \phi_{n_2}(\vec{R})
			\phi_{n_3}^*(\vec{R}) \phi_{n_4}(\vec{R}) 
			a_{n_1}^\dagger(t_1) a_{n_2}(t_1)
			a_{n_3}^\dagger(t_2) a_{n_4}(t_2)
		:\rangle \\
		=
	\frac{\eta^2}{4\pi^2}
	  \int\!\!\!\int\!\!\!\int\!\!\!\int \mathrm{d}\omega_{1,\ldots,4}
		\int\!\!\!\int_0^T\mathrm{d}t_{1,2}
		\sum_{n_1,\ldots,n_4} \phi_{n_1}^*(\vec{R}) \phi_{n_2}(\vec{R})
			\phi_{n_3}^*(\vec{R}) \phi_{n_4}(\vec{R})
		\langle a^{\dagger}_{n_1}(\omega_1)a^{\dagger}_{n_3}(\omega_3)
		a_{n_2}(\omega_2) a_{n_4}(\omega_4)\rangle \\
			\times
		\mathrm{e}^{\mathrm{i}\bigl[(\omega_1-\omega_2)t_1
		+(\omega_3-\omega_4)t_2\bigr]}
		\;.
\end{multline}	
Taking the average over the quantum mechanical expectation operators gives
nonzero contributions for three cases of the indices $n$: $n_1=n_2=n_3=n_4$,
$n_1=n_2\ne n_3=n_4$, and $n_1=n_4\ne n_2=n_3$, thus
 \begin{multline}
	n^{(2)}= \frac{\eta^2}{4\pi^2} \int\!\!\!\int_0^T\mathrm{d}t_{1,2}
		\int\mathrm{d}\omega \sum_n \betrag{\phi_n(\vec{R})}^4
		\langle [a^{\dagger}_{n}(\omega)]^2
		[a_{n}(\omega)]^2
		\rangle  \\
		+\frac{\eta^2}{4\pi^2}\int\!\!\!\int_0^T\mathrm{d}t_{1,2}
		\int\mathrm{d}\omega_{1,3} \sum_{n_1\ne n_3} 
		\betrag{\phi_{n_1}(\vec{R})}^2\betrag{\phi_{n_3}(\vec{R})}^2
		 \langle a^{\dagger}_{n_1}(\omega_1)
		a_{n_1}(\omega_1)a^{\dagger}_{n_3}(\omega_3)a_{n_3}(\omega_3)
		\rangle 
		 \\
	+\frac{\eta^2}{4\pi^2}\int\!\!\!\int \mathrm{d}\omega_{1,2}
		\int\!\!\!\int_0^T\mathrm{d}t_{1,2}
		\sum_{n_1\ne n_2} 
		\betrag{\phi_{n_1}(\vec{R})}^2\betrag{\phi_{n_2}(\vec{R})}^2	
			\langle a^{\dagger}_{n_1}(\omega_1)
		a_{n_1}(\omega_1)a^{\dagger}_{n_2}(\omega_2)a_{n_2}(\omega_2)
		\rangle 
		\mathrm{e}^{\mathrm{i}\bigl[(\omega_1-\omega_2)t_1
		+(\omega_2-\omega_1)t_2\bigr]} 
	\;.
	\label{eqLang1}
\end{multline}	
\end{widetext}	
This equation can be simplified by inserting the expectation values
from Eq.~(\ref{eqErwart1}). The double integral over $t_1$ and $t_2$
in the last line of Eq.~(\ref{eqLang1})
can be reduced to a single integral via
\begin{equation}
	\int\!\!\!\int_0^T\mathrm{d}t_{1,2}
	f(t_1-t_2) = \int_{-T}^T \mathrm{d}t f(t) [T-\betrag{t}]\;.
	\label{eqTtransform}
\end{equation}
The remaining two integrations over $\omega_1$ and $\omega_2$ in that line
each yield the Fourier transform $\hat{G}(t)$ of $G(\omega)$,
\begin{equation}
	\hat{G}(t)=\frac{1}{\sqrt{2\pi}}\int G(\omega)\mathrm{e}^{-\mathrm{i}\omega t}
		\mathrm{d}t\;.
\end{equation}
This thus gives
\begin{multline}
	n^{(2)}= \frac{\eta^2 T^2}{4\pi^2} 
		\int\mathrm{d}\omega \sum_n \betrag{\phi_n(\vec{R})}^4
		 \langle [a^{\dagger}_{n}(\omega)]^2
		[a_{n}(\omega)]^2
		\rangle  \\
		+\eta^2 T^2  \sum_{n_1\ne n_3}
		\betrag{\phi_{n_1}(\vec{R})}^2\betrag{\phi_{n_3}(\vec{R})}^2
		I_{n_1} I_{n_3} \\
	+\frac{\eta^2}{2\pi}
		\int_{-T}^T\mathrm{d}t
		\sum_{n_1\ne n_2} 
		\betrag{\phi_{n_1}(\vec{R})}^2\betrag{\phi_{n_2}(\vec{R})}^2	
		\\
		\times
			I_{n_1} I_{n_2} \hat{G}_{n_1}(t) \hat{G}^*_{n_2}(t)
		 [T-\betrag{t}]
	\;.
	\label{eqLang4}
\end{multline}	
The summation over $n_1\ne n_3$ is rather inconvenient. Thus, we include the
missing terms $n_1=n_3$ in that summation, yielding the value $[n^{(1)}]^2$,
and correct for this by subtracting in the first summation the terms just added.
There now appears the quantity
\begin{equation}
f_n I_n^2:=		\frac{1}{4\pi^2}\int\mathrm{d}\omega 
		\langle [a^{\dagger}_{n}(\omega)]^2
		[a_{n}(\omega)]^2
		\rangle  
		- \langle a^{\dagger}_{n}(\omega) a_{n}(\omega) \rangle^2
		 \;,
		 \label{eqDefF}
\end{equation}
that quantifies the deviation of the radiation in the $n$-th mode from a coherent
state. For coherent radiation, $f_n=0$ since then
$\langle [a^{\dagger}_{n}(\omega)]^2
[a_{n}(\omega)]^2 \rangle=\langle a^{\dagger}_{n}(\omega) a_{n}(\omega)
\rangle^2$ whereas for any Gaussian light source, in particular any object
emitting thermal radiation, $\langle [a^{\dagger}_{n}(\omega)]^2
[a_{n}(\omega)]^2 \rangle=2 \langle a^{\dagger}_{n}(\omega) a_{n}(\omega)
\rangle^2$, and $f_n = 1$. For classical radiation, $f_n\ge 0$ whereas
 for certain nonclassical radiation $f_n<0$ is possible.

Collecting results, using Eq.~(\ref{eqDefVar}), gives the variance of the
photocount
\begin{multline}
\var n=\overline{n}+\eta^2 T^2 \sum_n \betrag{\phi_n(\vec{R})}^4 I_n^2
	f_n + \\
	\frac{\eta^2}{2\pi} 
		\sum_{n_1\ne n_2} 
		\betrag{\phi_{n_1}(\vec{R})}^2\betrag{\phi_{n_2}(\vec{R})}^2	
			I_{n_1} I_{n_2} \\
				\times
				\int_{-T}^T\mathrm{d}t
				\hat{G}_{n_1}(t) \hat{G}^*_{n_2}(t)
		 [T-\betrag{t}]
	\;.
	\label{eqLang3}
\end{multline}

Equation~(\ref{eqLang3}) constitutes a core result of this paper. It gives the
general expression for the noise that is intrinsic to any measurement of a
multi-mode multi-frequency radiation field over some time $T$. The first term 
is the well-known
shot noise term that appears naturally in the quantum treatment: 
electric field and photocount are quantum-optically related via the factorial 
moment, see Eq.~(\ref{eqFactorialMoment1}), whereas in classical optics the
relation is via the ``plain'' moment, with the difference between factorial
and plain moments being precisely the
shot-noise term.

The second term, quantified by $f_n$ from Eq.~(\ref{eqDefF}), marks the excess
noise when the radiation field is not in a coherent state. As mentioned above,
$f_n=0$ for coherent radiation, and $f_n=1$ for thermal radiation.
Even when $f_n$ is small, 
any nonzero $f_n$ will eventually become
the dominating effect for the photonoise as its contribution increases
quadratically with time, in contrast to all other contributions. The cross-over
point, where this term becomes larger than the shot-noise term, is 
easily estimated 
from
Eq.~(\ref{eqLang3}) as $\eta T f_n I_n \gtrsim 1$.
For any thermal radiation this regime thus is already entered
once the measurement time is long enough for
one photon per mode to be counted on average. $I_n$ can be computed from
the Bose-Einstein factor,
evaluated at the temperature $\theta$ of the source,
\begin{equation}
	b(\omega,\theta) = \frac{1}{\exp(\hbar\omega/k_{\mathrm{B}}\theta)-1}\;.
\end{equation}
This value depends on the ratio of frequency and temperature, and thus has
the same value $b\approx 0.004$ for both a light bulb emitting visible light 
and an EUV plasma source operating at up to $200\,000~\mathrm{K}$ to emit
radiation with $\lambda\approx 13~\mathrm{nm}$. 

The third and final contribution in Eq.~(\ref{eqLang3}) quantifies the beating
of different frequency components when the measurement time is finite. It is
sometimes referred to as dynamic laser speckle since it is most relevant for
measurement times $T$ not much larger than the coherence time $\tau$ of the
radiation,
\begin{equation}
	\tau = \int_{-\infty}^\infty \mathrm{d}t \betrag{\hat{G}(t)}^2 
		= \int_{-\infty}^\infty \mathrm{d}\omega \betrag{G(\omega)}^2
		\;,
	\label{eqDefTau}
\end{equation}
hence the name ``temporal degree of coherence'' frequently given to
$\hat{G}(t)$. However, the factor $[T-\betrag{t}]$ is also relevant, and it
would be wrong to replace the integral simply by $\int \exp(-t/\tau)^2
\mathrm{d}t$, as we will now demonstrate. The extra factor $[T-\betrag{t}]$ ,
which seems to be largely ignored in phenomenological literature, gives less
weight to the frequently surprisingly large tails of $\hat{G}(t)$.

\begin{figure}
\includegraphics[width=0.95\columnwidth]{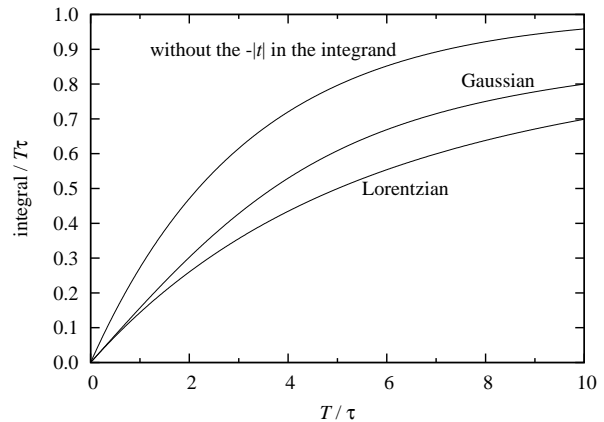}
\caption{Value of the integral over time in Eq.~(\protect\ref{eqLang3}) that
quantifies the strength of the dynamic speckle. The
two curves at the bottom follow by inserting the expressions for a Gaussian and
a Lorentzian spectrum from Eq.~(\protect\ref{eqKurvenformen}). When the term
$(T-\betrag{t})$ in Eq.~(\protect\ref{eqLang3}) is replaced by $T$, and an exponentially
decreasing $\hat{G}(t)$ is assumed, the third curve follows.}
\label{figKurvenformen}
\end{figure}

Two frequently encountered spectral distributions are Gaussian and Lorentzian,
\begin{align}
	G_{\mathrm{Gauss}}(\omega)&= \sqrt{2} \tau 
	\mathrm{e}^{-2\pi\tau^2\omega^2} \;,\\
	G_{\mathrm{Lorentz}}(\omega)&= \frac{2 \tau}{1 + 4\pi^2\tau^2\omega^2}
	\;.
\end{align}	
The curves above are already normalised to $1$, i.\,e., their integral (not
their square integral) is $1$. The widths of the spectral distributions have
been chosen such that the coherence time, computed from Eq.~(\ref{eqDefTau}), is
the same value $\tau$, allowing for a direct comparison of these two curves.
Their Fourier transforms  are
\begin{align}
	\hat{G}_{\mathrm{Gauss}}(t)&= \frac{1}{\sqrt{2\pi}} \exp\Bigl[
	-\frac{t^2}{8 \pi \tau^2} \Bigr]  \;,\\
	\hat{G}_{\mathrm{Lorentz}}(t) &= \frac{1}{\sqrt{2\pi}} \exp\Bigl[
		-\frac{\betrag{t}}{2\pi\tau} \Bigr] \;.
	\label{eqKurvenformen}
\end{align}	
The results for entering these expressions into the integral in
Eq.~(\ref{eqLang3}) are shown in Fig.~\ref{figKurvenformen}. There is a
significant dependence on the spectral shape, with the longer tails of the
Lorentzian leading to a slower approach to saturation. Please note that the
graph corresponds to the variance of the photocount. When the square root
of the variance is scaled
by the mean intensity -- this quantity is called the contrast --
the saturating curve shown in the figure turns into a curve decreasing to zero
as $T$ becomes larger.

$\hat{G}_{\mathrm{Lorentz}}$ is a plain exponential and thus conforms best to
the simple model of how to model finite temporal coherence. We thus have used
this curve for contrasting Eq.~(\ref{eqLang3}) by an approximation where the
factor $T-\betrag{t}$ is replaced by $T$, thereby ignoring the additional 
dependence on $\betrag{t}$. For all values of $T$, the integral increases by about
$40\,\%$, making the factor $T-\betrag{t}$ important for correct modelling of
the photon counting statistics.

A final word about the cross-over from ``dynamics'' to ``statics'' seems to be
in order. Since the temporal effect begins to saturate at time $\tau$ it is
frequently assumed that the nontemporal regime is entered once $T\gg\tau$. While
this statement is correct for single-mode radiation fields, it is incorrect here
since the shot-noise term $\overline{n}$, setting the reference value,  scales
linearly with the number $N$ of modes whereas the prefactor of the temporal term
scales as $N\cdot(N-1)$ such that the condition $T\gg\tau$ has to be replaced by
$T\gg N\tau$. Incidentally, since it is the regime $N\tau\gg T\gg\tau$ that
current high-power excimer lasers are operating in~\cite{noordman:09a}, this
distinction  is of actual importance.

\section{Etendue}
\label{secEtendue}

Similar to classical mechanics, theoretical optics knows the concept of phase
space~\cite{dragoman:02a}. The optical phase space is spanned by position and
spatial frequency, informally called $k$-vector, and the radiation field is
completely described by the (pseudo)-density function $W(\vec{r},\vec{k})$ known
as the Wigner function. Any radiation field then occupies a certain volume in
phase space. Unfortunately, there is no good
mathematical metric to actually measure or at least define such a volume.

In experimental and technological areas of optics, in contrast, the volume of
phase space occupied is well-defined and is referred to as etendue. The energy
density $I(\vec{r},\vec{\alpha})$ of the light, given as a function of position
and angle, is assumed to be larger than zero only inside a finite and
well-defined region. This allows for an easy and direct definition of the
etendue of the radiation field. Etendue is an important quantity since, at least
within geometrical optics, it can only be changed by vignetting the radiation
field. The etendue supplied by some radiation field incident on some optical
apparatus should thus be no larger than the etendue that this apparatus can
accept.

Convolving the Wigner function $W(\vec{r},\vec{k})$ with a
small kernel, its size given by an uncertainty-relation~\cite{schleich:01},
yields $I(\vec{r},\vec{\alpha})$ when the relation
\begin{equation}
	\alpha = k\lambda\;,
\end{equation}
between angle and $k$-vector is utilised. Thus on first sight it might seem that
the relation between microscopic and macroscopic definition would be
straight-forward.

The essential difference is that $W(\vec{r},\vec{k})$ has no finite support
whereas $I(\vec{r},\vec{\alpha})$ is assumed to have. It can be shown that due
to the uncertainty principle, every mode $\phi_n$ needs to have infinite tails
in at least either real space (position) or $k$-space
(angle)~\cite{folland:97a}, and this is reflected in $W(\vec{r},\vec{k})$. Of
course, also $I(\vec{r},\vec{\alpha})$ has the same infinite tails but in the
macroscopic world, the function values inside those tails drop down so quickly
that they are ``assumed'' to be zero once they are so low that they have become
irrelevant for practical purposes.

We thus need a definition for phase space volume that is based on a microscopic
radiation field but reduces to the macroscopic etendue for ``large'' radiation
fields. For the purpose of this paper, the relevant way of quantifying phase
space volume is by the number of modes that can be fit into it. We will tackle
the problem in the opposite way: We pick a certain number of ``sensible'' base
functions and compute all possible radiation fields that can be formed from
them. We will then identify for the intensity profiles $I(\vec{r})$
as function of position and $I(\vec{\alpha})$ as a function of direction 
the properties that
allow to deduce the number of base functions used.

To have shorter mathematical expressions, we will express directions in terms of
the $k$-vector instead of the angle $\alpha$. Furthermore we make two 
assumptions. First, we restrict ourselves to one spatial dimension but the
extension is straight-forward and becomes trivial when the radiation field
factorises in its spatial dimensions. Second, we assume that the intensity
distribution $I(x,k)$ factorises into a spatial term $I^{(x)}$ and an angular
term $I^{(k)}$. Since
the etendue accepted by practically any real-life apparatus factorises,
these assumptions do not limit the applicability of this approach.

We start with $N+1$ orthonormal base functions $\phi_n(x)$, $n=0,\ldots,N$.
The
orthogonality is essential since this prevents the base functions 
from being too similar and thus collapsing into a very small volume of phase space.
Furthermore, all modes of the radiation field have to be
mutually orthogonal anyhow as they are eigenfunctions of a Hermitian operator,
cf.~Eq.~(\ref{eqGamma2}). We assume them to be normalised for the ease of the
calculation. For every $\phi_n(x)$, there also exists its Fourier transform
\begin{equation}
	\hat{\phi}_n(k)=\frac{1}{\sqrt{2\pi}}\int_{-\infty}^\infty \phi_n(x)  
	\mathrm{e}^{-\mathrm{i}k x} \mathrm{d}x\;.
\end{equation}

To progress, we need to choose a certain set of base functions. The aim is
to pack the base functions as tightly as possible in phase space, and without loss
of generality we try to pack them around the origin. If all $\phi_n$,
$n=0,\ldots,N$, are assumed to fulfil the conditions
\begin{subequations}
\label{eqHardy1}
\begin{align}
	\betrag{\phi_n(x)}&\le C(1+\betrag{x})^N \mathrm{e}^{-a\betrag{x}^2}\;,\\
	\betrag{\hat{\phi}_n(k)}&\le C(1+\betrag{k})^N \mathrm{e}^{-b\betrag{k}^2}\;,
\end{align}	
\end{subequations}
then it can be shown~\cite{hardy:33a} that $a b=1/4$ is the most condensed
situation for which there are solutions. Furthermore, all solutions can then be
written as
\begin{equation}
	\phi_n(x)=P_N(x) \mathrm{e}^{-a\betrag{x}^2}\;,\qquad
	\hat{\phi}_n(k)=\tilde{P}_N(k) \mathrm{e}^{-b\betrag{k}^2}
	\label{eqHardy2}
\end{equation}
where $P_N(x)$, $\tilde{P}_N(k)$ are polynomials of order not higher than $N$. 
Only the product $a b$ is fixed by this but not $a$ or $b$ on their own. This
freedom is equivalent to changing $\phi_n(x)\to \phi_n(x/r)$ with the
simultaneous change $\hat{\phi}_n(k)\to \hat{\phi}_n(k r)$. This also
reflects that, while phase space volume is well-defined and preserved 
as volume, its projection onto a particular axis is allowed to change.

The set of solutions $\phi_n(x)$ of Eq.~(\ref{eqHardy2}) is $N+1$ dimensional,
and a convenient orthonormal set consists of the Hermite functions
$\psi_n(x)$, defined as
\begin{equation}
	\psi_n(x/r)=\frac{1}{\sqrt{n!2^n\sqrt{\pi}}}
		\exp\left[-\frac{x^2}{2r^2}\right] H_n(x/r)\;,
	\label{eqDefHermite1}
\end{equation}
where $H_n(x)$ are the Hermite polynomials,
\begin{multline}
	H_0(x)=1\;,\qquad H_1(x)=2x\;, \\
	H_{n+1}(x)=2 x H_n(x)-2 n H_{n-1}(x)\;.
\end{multline}
The prefactor in Eq.~(\ref{eqDefHermite1}) ensures normalisation of $\psi_n(x)$. 
The Hermite functions are eigenfunctions of the Fourier transform,
\begin{equation}
	\hat{\psi}_n( \xi r) = (-\mathrm{i})^n \psi_n(\xi / r )\;,
\end{equation}
such that the derivation presented in the following for real space
also applies to $k$-space.

All normalised functions $\phi(x)$ inside the solution space of 
Eq.~(\ref{eqHardy1}) can be written as
\begin{equation}
	\phi(x)=\sum_{n=0}^N a_n \psi_n(x)\;,\qquad
	\sum_{n=0}^N \betrag{a_n}^2=1 \;.
\end{equation}
The maximum intensity $\betrag{\phi(x)}^2$ possible for such a state can
be computed using the method of Lagrange multipliers. 
Finding an extremum of $\phi(x)$ is equivalent to finding one of
$\betrag{\phi(x)}^2$, and we will pick either of these quantities depending on
which one is more convenient.
The condition $\partial\Lambda/\partial a_0=\ldots=
\partial\Lambda/\partial a_N=\partial\Lambda/\partial\mu=0$ of the function
\begin{equation}
	\Lambda(a_0,\ldots,a_N,\mu)=\sum_{n=0}^N a_n \psi_n(x)-
			\mu( 1 - \sum_{n=0}^N \betrag{a_n}^2 ) \;,
\end{equation}
yields the intermediary result $\psi_k(x)+2\mu a_k=0$. By multiplying this
with $\psi_k(x)$ respectively $a_k$ and summing, this gives the two conditions
\begin{subequations}
\label{eqLagrange3}
\begin{align}
	\sum_{n=0}^N \betrag{\psi_n(x)}^2 &= - 2 \mu \sum_{n=0}^N a_n \psi_n(x)
		= -2 \mu \phi(x) \;, \\
- 2 \mu \sum_{n=0}^N \betrag{a_n}^2 &=
	\sum_{n=0}^N a_n \psi_n(x) = \phi(x) \;.
	\label{eqLagrange3b}
\end{align}
\end{subequations}
Solving for $\mu$ by eliminating $\phi(x)$ and inserting $\mu$ into
Eq.~(\ref{eqLagrange3b}), remembering that $\sum_{n=0}^N \betrag{a_n}^2=1$,
one arrives at the result for the maximum intensity, namely
\begin{equation}
	\max \betrag{\phi(x)}^2=\sum_{n=0}^N \betrag{\psi_n(x)}^2\;.
	\label{eqLagrange4}
\end{equation}
The method of Lagrange multipliers only gives necessary but not sufficient 
conditions for extrema, meaning that there cannot be any additional extrema
different from Eq.~(\ref{eqLagrange4}) but this equation might not describe
an extremum in the first place.
Equation~(\ref{eqLagrange4}) actually describes
two solutions, $\phi(x)=+\ldots$ and $\phi(x)=-\ldots$.
It is obvious from physics reasons that at least
two extrema, one being the most positive and one the most negative,
exist. Hence, Eq.~(\ref{eqLagrange4}) uniquely describes these two
extrema.

\begin{figure}
\includegraphics[width=0.85\columnwidth]{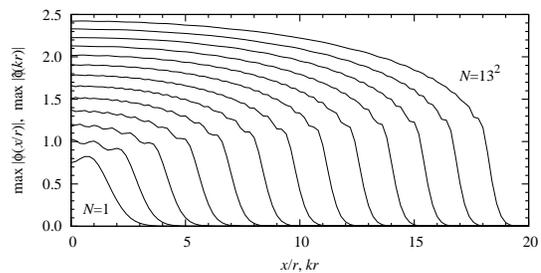}
\caption{Maximum values $\max\betrag{\phi(x)}$ and
$\max\betrag{\hat{\phi}(k)}$, computed from Eq.~(\ref{eqLagrange4}), as a
function of $N$ where $N$ is the number of basis functions minus 1. 
Depicted are the curves for $N=1,4,9,\ldots,13^2$. If the
corresponding curves for a normalised radiation field are below a given line,
the radiation field cannot have more modes than indicated by
the number $N$ on the curve.
}
\label{figLagrange1}
\end{figure}

A graphical display of the solutions $\max\betrag{\phi(x)}$ and
$\max\betrag{\hat{\phi}(x)}$ can be found in Fig.~\ref{figLagrange1} as a
function of $N$. The right edge of $\max\betrag{\phi(x)}$ becomes progressively
steeper as $N$ increases. This means that in the macroscopic limit $N\gg 1$
there exist a sharp edge that can be associated with the macroscopic etendue. To
define the position of this edge,  one could pick the $x$-coordinate where
$\max\betrag{\phi(x)}^2=1/2$ or where it is equal some other particular value
but this would lead to an arbitrary result. It can be shown that the number of
basis functions is then proportional to the  square of the width at which the
``cut'' is done~\cite{powell:05a} such that the result depends more on this
arbitrary decision than on the radiation field. Also the method of ``almost
bandwidth-limited functions'' (which is usually formulated  for time-dependent
electric signals, hence the term ``bandwidth'' instead of  ``angular range'')
suffers from a similar problem~\cite{slepian:83a}.

Luckily, there exists a strong link between the microscopic and 
macroscopic
worlds because the Hermite functions $\psi_n(x)$ are the solutions of the
eigenvalue equation
\begin{equation}
	\frac{\mathrm{d}^2\psi_n(x)}{\mathrm{d}x^2}+(2n+1-x^2)\psi_n(x)=0\;,
	\label{eqHarmonic1}
\end{equation}
that can be found in any quantum mechanics textbook as it describes the harmonic
oscillator. The eigenfunction $\psi_n$ has eigenvalue (=energy) of $n+1/2$, and
a classical oscillator with the same energy is restricted to the interval
$-\sqrt{2n+1}\le x\le \sqrt{2n+1}$. The extremal curve $\phi(x)$ computed above
thus is the extremal superposition of harmonic-oscillator solutions with energy
level $n$ up to $N$. Hence, the classical edge beyond which the oscillator
cannot be found is given by $\sqrt{2N+1}$.

Summarising, a microscopic radiation field has at most $N+1$ modes when for all of its
modes $\phi_m(x)$ the conditions
\begin{subequations}
\begin{align}
	\betrag{\phi_m(x)}^2 & \le \sum_{n=0}^N \betrag{\psi_n(x/r)}^2 \;,\\
	\betrag{\hat{\phi}_m(k)}^2 & \le \sum_{n=0}^N 
		\betrag{\psi_n(k r)}^2 \;,
\end{align}
\end{subequations}
are fulfilled for some value $r$ identical for all modes. Since
$\betrag{\phi_m(x)}^2$ and $\betrag{\hat{\phi}_m(k)}^2$ are  the intensities as
a function of position and angle, respectively, these are quantities that can in
principle be measured.

A macroscopic radiation field with a half-diameter $X$ in real space
and a half-diameter $K$ in angle space is thus spanned by modes $\psi_n$ with
$n=0,\ldots,N$ determined by
\begin{equation}
	r\sqrt{2N+1}=X\;,\qquad \frac{1}{r}\sqrt{2N+1}=K\;.
\end{equation}
The scale factor $r$ drops out when expressing results in terms of the
macroscopic etendue.
For given etendue $E\equiv XK$, the number $N$ of modes (minus 1) is thus
\begin{equation}
	N=\frac{E-1}{2} \;.	
\end{equation}
If the etendue is specified in angle (radians) instead of a $k$-vector,
this becomes
\begin{equation}
	N=\frac{E/\lambda-1}{2} \;.
\end{equation}

\section{Static speckle}
\label{secStaticSpeckle}

The variance of the photocount computed in Sec.~\ref{secZeitunschaerfe}
describes the uncertainty of the photocount if the radiation field
at the detector is completely known. Frequently,
the mode structure is known at the light source but not at the
detector. The standard undergraduate example is a HeNe-laser pointed at the wall
of the lecture room where the observer can see a speckle pattern. When the eyes
are then moved, the direction of motion of the observed speckle pattern depends
on whether the viewer is near-sighted or far-sighted~\cite{goodman:07}. The
HeNe-laser emits a very regular mode structure but the modes are then changed by
scattering at the rough wall. The influence of the eye movement demonstrates
that scattering at the wall does not result in local intensity changes at the
place of scattering but
rather in a deformation of the mode structure that translates into intensity
changes only by propagation to and focusing in the eye.

The change of the mode structure on its way from the radiation source to the
detector can be unintentional, just as in the example above, or it can be
intentional due to some apparatus designed to mix the modes of the radiation
field. Unless optical systems are manufactured to the highest standards, they
will have surface roughness in excess of the wavelengths. This results in
uncontrolled -- thus in a certain sense random --  mode deformations. In
addition, any
apparatus designed to provide a uniform and stable illumination at its exit can
only do this by means of ``mixing'' the incoming light. Illumination systems
used in optical lithography are the supreme example of such an apparatus as the
local intensity at the output varies only by about $1\,\%$ even when the form of
the illumination of its entrance is completely changed. While the light paths
inside such an apparatus might be very controlled, for an outside observer there
(intentionally) is no recognisable connection between the light at the entrance
and the exit. In other words, an apparently random change of the mode structure
occurs in this case as well.

In Sec.~\ref{secZeitunschaerfe} we have used the modes $\phi_n(\vec{r})$ of the
radiation field to expand the operator for the electric field,
cf.~Eq.~(\ref{eqFexpansion}). Any orthonormal basis could have been used instead
but using the modes offered the advantage that the associated annihilation
operators $a_n(\omega)$ become uncorrelated, cf.~Eq.~(\ref{eqErwart1}). We will 
now switch to a different basis set, at the expense of replacing
Eq.~(\ref{eqErwart1}) by something more complicated.

\begin{figure}
\includegraphics[width=0.8\columnwidth]{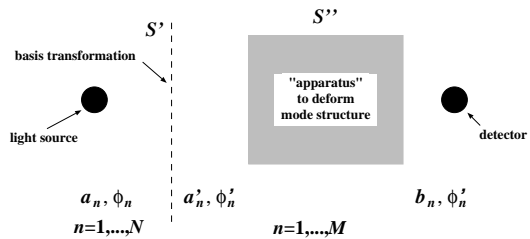}
\caption{Graphical display of the transformation steps to arrive at static
speckle. By means of a basis transformation, described by $S'$, one switches
from the modes $\phi_n$ that are optimal to describe the light source, to new
base functions $\phi'_n$ that are optimal to describe the photodetector.
Along with the transformation $\phi\to \phi'$, the annihilation operators
 also need to be
transformed, $a\to a'$. The additional action of scattering or mixing is then
described by the matrix $S''$, transforming $a'$ into $b$.}
\label{figSkizzeStatisch}
\end{figure}

Considering only the radiation source and the detector,
hence ignoring the scattering between them for the moment,
we pick new
basis functions $\phi'_n(\vec{r})$ and associated annihilation operators $a'_n$
such that at the location $\vec{R}$ of the detector all $\phi'_n$ except for
$\phi'_1$ vanish. Such a basis set can always be found and offers the advantage that
photodetection at the point $\vec{R}$ has been reduced to the photocount of the
operator $a'_1$. 
Since both bases are orthonormal, the transformation between 
$\vec{a}=a_1,a_2,\ldots$ 
and $\vec{a}'=a'_1,a'_2,\ldots$ can be described by a unitary matrix $S'$,
\begin{equation}
	\vec{a}'= S'\vec{a}\;,
	\label{secSmatrix}
\end{equation}
with $S S^\dagger=\openone$. 
Computing $S'$ explicitly would be a very formidable task 
but it will turn out that knowledge of $S'$ is not needed. 

We will fix the number $M$ of new basis functions $\phi'_n$ shortly but need to
allow for the case that $M$ is larger than the number $N$ of excited modes
of the radiation field. This is easily achieved by simply adding $M-N$
vacuum states to the input for Eq.~(\ref{secSmatrix}).

We now assume that the radiation field is perturbed between the light source
and the place of the detector by means of some ``virtual device'' -- either
intentionally by a properly designed apparatus or, usually unintentionally,
by quasi-random scattering. This can be
described in two opposite but physically equivalent approaches. Either, one
assumes that the mode structure is perturbed, thus changing $\{\phi'_n\}$ while
keeping $\{a'_n\}$ unchanged, or one assumes that the mode structure
$\{\phi'_n\}$ is unchanged while energy is transferred between modes, thus
modifying $\{a'_n\}$.
We use the second approach, also known as
the method of input-output relations~\cite{jeffers:93a},
since the annihilation operators $\{b_n\}$ describing the radiation
leaving the device can then be expressed in terms of the annihilation operators
$\{a'_n\}$ entering device. In the absence of nonlinear media, this relation
is linear, and can be described by a matrix $S''$
\begin{equation}
	\vec{b}= S''\vec{a}'\;.
	\label{secSmatrix2}
\end{equation}
If there is no absorption, $S''$ is unitary. Apart from energy conservation,
already the commutation relation~(\ref{eqVertauschungsrelation}) demands the unitarity
of $S$.
In the presence
of absorption, Eq.~(\ref{secSmatrix2}) would need to be supplemented by
noise sources~\cite{jeffers:93a} to ensure these commutation relations. 

In the absence of further information, the unitary matrix $S'$ describing the 
distortion of the mode structure is uniformly distributed in the space of
unitary matrices~\cite{beenakker:97a}. This concept of ``maximum uncertainty''
is known from other areas of physics as well, most notably from
thermodynamics where the basic lemma is that every microstate compatible with
the macroscopic boundary conditions is equally likely.

The photocount distribution at
some point $\vec{R}$ has thus been reduced to the photocount distribution for
the mode $b_1$, related to the original incident mode structure 
of the radiation source by
\begin{equation}
	\vec{b} = S \vec{a}\;,\qquad S=S' S''\;.
	\label{eqSstatischGesamt}
\end{equation}
Since for a given position of detector and light source $\vec{R}$, $S'$ 
is a fixed
unitary matrix, the product $S=S' S''$ is also distributed
uniformly~\cite{mehta:90}, and so no knowledge of $S'$ is needed, as promised
above.

The only information about what happens between light source and detector
enters in the form of the number $M$ of basis functions. This number is not
arbitrary but is determined by the physics of the mode deformation process
described by the matrix $S''$. Please remember that $M$ is the number of base
functions that are mixed by the transformation $S''$. 
If the mode deformation is due to some intentional mixing in an apparatus
accepting and emitting some etendue $E$, the number $M$ follows from the results
of Sec.~\ref{secEtendue} and has a finite, well-defined value. If the mode
deformation is due to scattering or a diffuser, $M$ is determined by the number
of basis functions that could in principle end up in the detector due to
scattering or diffusing. Depending on maximum scattering angle, this can be a
very large number and is always larger than $N$.

Computing the first factorial moment analogue to Eq.~(\ref{eqAfrequenz2}) 
yields
\begin{multline}
n^{(1)}=
\eta \langle \int_0^T \mathrm{d}t
	b^{\dagger}_1(t) b_1(t) \rangle
	\\ 
	= \eta \langle
	\int_0^T \mathrm{d}t \sum_{nm} a^{\dagger}_n(t) S^{\dagger}_{n1}
	S_{1m} a_m(t) \rangle = \eta T \sum_n \langle S_{1n} S^{\dagger}_{n1}
	\rangle I_n
	\;,
	\label{eqAfrequenz3}
\end{multline}	 
where in the final equality sign all averages over the fluctuations of
electromagnetic field have been taken, just as described in
Sec.~\ref{secZeitunschaerfe}, and an average over the ensemble of random unitary
matrices remains. The average is easily computed from $S S^\dagger=\openone$,
hence the average in Eq.~(\ref{eqAfrequenz3}) is equal to $1/M$ where $M$ is the
order of the matrix, yielding
\begin{equation}
	n^{(1)}=\frac{\eta T}{M} \sum_n I_n\;.
\end{equation}	
This is the expected result, namely that the mean intensity 
at the detector is equal to the properly scaled total intensity of
the original light field.

Computing the first moment demonstrated that to translate the equations
from Sec.~\ref{secZeitunschaerfe}, all terms of the form $\phi_n(\vec{R})$
have to be replaced by $S_{1n}$, and all terms $\phi^*_n(\vec{R})$
by $S^{\dagger}_{n1}$. The correct starting point is Eq.~(\ref{eqLang4}), which
transforms into
\begin{multline}
	n^{(2)}= \frac{\eta^2 T^2 }{4\pi^2}
		\int\mathrm{d}\omega \sum_n \langle \betrag{S_{1n}}^4 \rangle
		 \langle [a^{\dagger}_{n}(\omega)]^2
		[a_{n}(\omega)]^2
		\rangle  \\
		+\eta^2 T^2  \sum_{n_1\ne n_3}
		\langle \betrag{S_{1n_1}}^2 \betrag{S_{1n_3}}^2 \rangle
		I_{n_1} I_{n_3} \\
	+\frac{\eta^2}{2\pi}
		\int_{-T}^T\mathrm{d}t
		\sum_{n_1\ne n_2} 
		\langle \betrag{S_{1n_1}}^2 \betrag{S_{1n_2}}^2 \rangle	
		\\
		\times
			I_{n_1} I_{n_2} \hat{G}_{n_1}(t) \hat{G}^*_{n_2}(t)
		 [T-\betrag{t}]
	\;.
	\label{eqLang6}
\end{multline}	
It only remains to compute the necessary averages over the unitary group in a 
similar spirit
to above where we found that $\langle S_{1n}S^{\dagger}_{n1}\rangle=1/M$.
The average of the square of this expression 
is frequently needed and can be found in many
papers, e.\,g. in Ref.~\onlinecite{brouwer:96a}, with the result
\begin{equation}
	\langle [S_{1n}S^{\dagger}_{n1}]^2 \rangle = \frac{2}{M(M+1)} \;.
	\label{eqSquadratMittel}
\end{equation}
The average $\langle \betrag{S_{1n}}^2 \betrag{S_{1m}}^2 \rangle$ for
$n\ne m$ is not identical to $\langle \betrag{S_{1n}}^2 \rangle 
\langle \betrag{S_{1m}}^2 \rangle$ as the unitary condition $S
S^\dagger=\openone$ introduces correlations among the elements of $S$. 
Starting from the exact
relation $1=\sum_{nm} \betrag{S_{1m}}^2\betrag{S_{1n}}^2$ allows us to write
\begin{equation}
	\sum_{n\ne m} \langle \betrag{S_{1m}}^2\betrag{S_{1n}}^2 \rangle
	+ \sum_n \langle \betrag{S_{1n}}^4 \rangle = 1\;,
\end{equation}
which, together with Eq.~(\ref{eqSquadratMittel}), gives us the desired 
average,
\begin{equation}
	\langle \betrag{S_{1n}}^2 \betrag{S_{1m}}^2 \rangle
	= \begin{cases}
		\frac{1}{M(M+1)}	& n\ne m \\
		\frac{2}{M(M+1)} 	& n=m
	\end{cases}
	\label{eqMittelwerteS}
\end{equation}	
This allows us to transform Eq.~(\ref{eqLang6}) into
\begin{multline}
	n^{(2)}= \frac{\eta^2 T^2 }{2\pi^2 M(M+1)}
		\int\mathrm{d}\omega \sum_n 
		 \langle [a^{\dagger}_{n}(\omega)]^2
		[a_{n}(\omega)]^2
		\rangle  \\
		+\frac{\eta^2 T^2}{M(M+1)}  \sum_{n_1\ne n_3}
		I_{n_1} I_{n_3} \\
	+
		\frac{\eta^2}{2\pi M(M+1)}
		\int_{-T}^T\mathrm{d}t
		\sum_{n_1\ne n_2} 
			I_{n_1} I_{n_2} \hat{G}_{n_1}(t) \hat{G}^*_{n_2}(t)
		 [T-\betrag{t}]
	\;.
	\label{eqTemp1exakt}
\end{multline}
When computing the variance in Sec.~\ref{secZeitunschaerfe}, the term
with 
$I_{n_1} I_{n_3}$ dropped out when $[n^{(1)}]^2$ was subtracted from
$n^{(2)}$. This is no longer the case here, and one arrives at
\begin{multline}
	\var n= \overline{n} 	
		- \frac{1}{(M+1)} \overline{n}^2
		+
		\frac{\eta^2 T^2}{M(M+1)} \sum_n ( 2 f_n + 1 ) I_n^2
		\\
	+
		\frac{\eta^2}{2\pi M(M+1)}
		\int_{-T}^T\mathrm{d}t
		\sum_{n_1\ne n_2} 
			I_{n_1} I_{n_2} \hat{G}_{n_1}(t) \hat{G}^*_{n_2}(t)
		 [T-\betrag{t}]
	\;.
	\label{eqTemp2exakt}
\end{multline}

Before discussing this result, we will analyse an approximation that
is frequently done: The exact averages from
Eq.~(\ref{eqMittelwerteS}) are replaced by their Gaussian approximation,
equivalent to a phasor representation~\cite{goodman:07},
\begin{equation}
	\langle \betrag{S_{1n}}^2 \betrag{S_{1m}}^2 \rangle
	\approx \begin{cases}
		\frac{1}{M^2}	& n\ne m \\
		\frac{2}{M^2} 	& n=m
	\end{cases}
	\label{eqMittelwerteSnaeherung}
\end{equation}
The denominator in Eq.~(\ref{eqMittelwerteS}) was $M(M+1)$ so that the mixing
of the modes would conserve energy. Equation~(\ref{eqMittelwerteSnaeherung})
neglects this constraint. While this error might seem to be negligible for large
$M$, we will demonstrate that its effect is surprisingly large. 
First, we compute
\begin{multline}
	n^{(2)}\approx \frac{\eta^2 T^2 }{2\pi^2 M^2}
		\int\mathrm{d}\omega \sum_n 
		 \langle [a^{\dagger}_{n}(\omega)]^2
		[a_{n}(\omega)]^2
		\rangle  \\
		+\frac{\eta^2 T^2}{M^2}  \sum_{n_1\ne n_3}
		I_{n_1} I_{n_3} \\
	+
		\frac{\eta^2}{2\pi M^2}
		\int_{-T}^T\mathrm{d}t
		\sum_{n_1\ne n_2} 
			I_{n_1} I_{n_2} \hat{G}_{n_1}(t) \hat{G}^*_{n_2}(t)
		 [T-\betrag{t}]
	\;.
\end{multline}
In contrast to Eq.~(\ref{eqTemp1exakt}), the term with $I_{n_1} I_{n_2}$ 
now cancels against $[n^{(1)}]^2$. The variance then becomes
\begin{multline}
	\var n\approx \overline{n} +	
		\frac{\eta^2 T^2}{M^2} \sum_n ( f_n + 1 ) I_n^2
		\\
	+
		\frac{\eta^2}{2\pi M^2}
		\int_{-T}^T\mathrm{d}t
		\sum_{n_1\ne n_2} 
			I_{n_1} I_{n_2} \hat{G}_{n_1}(t) \hat{G}^*_{n_2}(t)
		 [T-\betrag{t}]
	\;.
\end{multline}
This is basically Eq.~(\ref{eqLang3}) with an additional contribution
\begin{equation}
	\var n_{\mathrm{static}} \approx \frac{\eta^2 T^2}{M^2} \sum_n I_n^2\;,
	\label{eqVarNstatisch}
\end{equation}
that is called ``static'' as its scaling with $T^2$ implies that its strength,
when scaled by the mean squared intensity, is independent of measurement time
$T$. 

We arrived at the ensemble of random $S$ for calculating the
variance $\var n_{\mathrm{static}}$  by
assuming a fixed position of the detector while the mode deformations
between radiation source and detector is different
for a every member of the ensemble.
However, from Eq.~(\ref{eqSstatischGesamt}) it follows that this variance
can equally well be computed or measured by keeping the mode formations
constant while changing the position of the detector. The static speckle
contrast $c$,
\begin{equation}
	c\equiv \frac{\sqrt{\var n_{\mathrm{static}}}}{
		\overline{n}} \approx 
		\frac{\bigl[\sum_n I_n^2]^{1/2}}{\sum_n I_n} \;,
\end{equation}
can thus equally well be determined by doing a (long) measurement at different detector 
positions $\vec{R}$.
In practise, one uses a camera instead of a point-like
photodetector, and $c$ is calculated from the observed  contrast of the image
taken by the camera. The image resembles the speckle pattern found on the fur or
skin of many animals, hence the name ``speckle''. The static speckle contrast
can, in this approximation, conveniently be expressed in terms of the  effective
number of modes $N_{\mathrm{eff}}$,
\begin{equation}
	N_{\mathrm{eff}} = \frac{\bigl[\sum_n I_n]^2}{\sum_n I_n^2} \;,
\end{equation}
that is equal to the actual number of modes if all intensities are equal, and
smaller (implying more static speckle) otherwise.	

For the exact result~(\ref{eqTemp2exakt}), static speckle is not so easy to
define, as the difference between Eq.~(\ref{eqTemp2exakt}) and the results from
Sec.~\ref{secZeitunschaerfe} is not a simple term proportional to $T^2$.
However, taking the limit $T\to\infty$, hence living up to the name ``static'',
the difference between these two terms gives
\begin{align}
	c&=\frac{\bigl[ \var n_{\mathrm{Eq.\,(\ref{eqTemp2exakt})}}
		- \var n_{\mathrm{Eq.\,(\ref{eqLang3})}}\bigr]^{1/2}}{
		\overline{n}} \nonumber\\
	&= \left[ \frac{M}{M+1} \frac{\sum_n I_n^2}{\bigl[\sum_n I_n\bigr]^2} 
	+ \frac{M-1}{M+1} \frac{\sum_n f_n I_n^2}{\bigl[\sum_n I_n\bigr]^2}
	-\frac{1}{M+1}\right]^{1/2} \nonumber\\
	&= \left[ \frac{M}{M+1} \Bigl(\frac{1}{ N_{\mathrm{eff}}}
		-\frac{1}{M} \Bigr)
	+ \frac{M-1}{M+1} \frac{\sum_n f_n I_n^2}{\bigl[\sum_n I_n\bigr]^2}
	\right]^{1/2} \;.
	\label{eqCexakt}
\end{align}		

There thus is a very significant difference between the exact result and the
approximation. In the latter, the contribution~(\ref{eqVarNstatisch})  is
minimal but still nonzero if all modes carry the same intensity $I_n$ whereas
for the exact result Eq.~(\ref{eqTemp2exakt}), assuming coherent radiation, this 
contribution vanishes if all $I_n$
are identical since then $N_{\mathrm{eff}}=M$. 

The approximation becomes valid in the limit that there are many modes but only
a small number of them are actually excited: Equation~(\ref{eqVarNstatisch})
depends on the average squared intensity $\langle I_n^2\rangle_n$ of the modes
of the light source, where for a fixed light source the average is taken over
its modes. This kind of average is indicated by the notation $\langle\ldots
\rangle_n$. In contrast, the exact result Eq.~(\ref{eqTemp2exakt}) depends on
$M/(M+1)[\langle I_n^2\rangle_n-\langle I_n\rangle_n^2]$ but the term $\langle
I_n\rangle_n^2$ scales with the square of the fraction of excited modes and may
thus be neglected in the limit stated above.

This important conclusion can also be formulated in terms of the etendue
introduced in Sec.~\ref{secEtendue}. The approximation is valid in the
limit that the light source fills only a small fraction of the etendue that the
``device'' used for mixing can accept. It also means that mixing will not
introduce static speckle as long as the entire etendue of the mixer is
uniformly filled by the light source (assuming coherent radiation).

This might seem to contradict phenomenological theories that claim that there
should aways be some pattern in the detector plane because light from the same
point of the light source can reach the detector along different paths, thereby
resulting in uncontrolled interference that inevitably creates intensity
fluctuations. However, the unitary condition $S S^\dagger=\openone$ ensures
that, whenever a mode of the light source is brighter at the detector, another
mode has to be dimmer. This property is fulfilled for any individual $S$ since
$\sum_n \betrag{S_{1n}}^2=1$, and no averaging over $S$ is necessary to arrive
at this conclusion.

Technologically, mixing apparatuses have a well-defined etendue, hence a
well-defined value of $M$. When the etendue that they can accept, is filled
to a large extend by the incident radiation, the effect of the unitary
condition is strong, and the exact formula~(\ref{eqTemp2exakt}) has to be used.
For random scattering, on the other hand, in particular by a diffuser,
frequently $N\ll M$, and therefore the
approximate formula~(\ref{eqVarNstatisch}) might suffice.

\section{Static speckle and etendue}
\label{secStaticKorrelation}

In Sec.~\ref{secStaticSpeckle} we demonstrated that the static speckle depends
on the mean squared intensity in each mode of the radiation source, hence on
$\sum_n I_n^2$, cf. Eqs.~(\ref{eqTemp2exakt}) and~(\ref{eqVarNstatisch}). The
effect is minimal if all modes carry equal weight, and the maximum number of
modes possible follows from the etendue via Sec.~\ref{secEtendue}. The minimum
amount of static speckle possible can then be computed -- it is zero if the
etendue is not increased by the scattering and mixing processes, and finite
otherwise. 

This raises the following question: is it possible not only to compute the
minimum amount of static speckle for given etendue of the radiation field but
also the ``typical'' amount of speckle? We will demonstrate that static speckle
contrast is a self-averaging quantity, i.\,e., in the limit of many modes $M$,
the relative fluctuation while taking the average over the unitary group becomes
small. This implies that the ensemble average is characteristic for almost all
realisation of the ensemble.

When $\phi_n(\vec{r})$ are the modes of the electric field, the electric field
$E(\vec{r})$ is uniquely described by coefficients $a_n$,
\begin{equation}
	E(\vec{r})=\sum_n a_n \phi_n(\vec{r})\;,
	\label{eqFeld1}
\end{equation}
and the static speckle contrast is determined by the effective number
$N_{\mathrm{eff}}$ of modes in a more or less complicated way depending on
whether the exact result or the approximation is used,
\begin{equation}
	\frac{1}{N_{\mathrm{eff}}}
	=\frac{\sum_n \betrag{a_n}^4}{
		\bigl[\sum_n \betrag{a_n}^2\bigr]^2} \;.
	\label{eqNeff1}
\end{equation}

In Sec.~\ref{secEtendue} the linear span corresponding to some given etendue $E$
was computed. In the absence of dynamic information, one cannot select the modes
out of all possible sets of basis vectors, and any set of basis vectors could
thus represent the modes. We assume that the
decomposition~(\ref{eqFeld1}) represents the actual modes $\phi_n(x)$ of the 
radiation field. Since, as just explained, we cannot know $\phi_n(x)$,
the equations for the
static speckle contrast are instead evaluated using some other
orthonormal basis $\phi'_n(\vec{r})$. The two bases are related by
a unitary transformation, $\vec{\phi}(\vec{r})= U \vec{\phi}'(\vec{r})$, and 
consequently $\vec{a}'=U\vec{a}$, but $U$ is unknown for the reasons just
explained.

If no additional information is available, the best available estimate
is the average of 
\begin{align}
	\frac{1}{N'_{\mathrm{eff}}}&= \frac{\sum_n \betrag{a'_n}^4}{
		\bigl[\sum_n \betrag{a'_n}^2\bigr]^2}
		\nonumber \\
	&=	\frac{\sum_{n,i_1,\ldots,i_4}
		U^\dagger_{n,i_1} U_{n,i_2} U^\dagger_{n,i_3} U_{n,i_4}
		a^*_{i_1}a_{i_2}a^*_{i_3}a_{i_4}
		}{
		\bigl[\sum_n \betrag{a_n}^2\bigr]^2} \;,
	\label{eqNeffMittel}
\end{align}
after integrating over the unitary group, 
i.\,e., over all possible unitary matrices. We study the average of 
$1/N'_{\mathrm{eff}}$ instead of $N'_{\mathrm{eff}}$
since it both is easier to compute and is the relevant quantity for the speckle
contrast. The variance $\langle [1/N'_{\mathrm{eff}}]^2\rangle
-\langle 1/N'_{\mathrm{eff}}\rangle^2$ then
specifies how well the average is representative for all possible members
of the ensemble.

Averaging over the unitary group, the nonvanishing terms 
are of the form
\begin{multline}
\langle U^*_{i_1,j_1} \cdots U^*_{i_M,j_M} U_{P(i_1),P'(j_1)}
	\cdots U_{P(i_M),P'(j_M)}\rangle \\
	= V_{s(P^{-1}P')} \;,
\end{multline}
where $P$ and $P'$ are permutations of the numbers $1,\ldots,M$,
and $s$ denotes the cycle structure of a permutation. The coefficients $V$ 
depend only on the cycle structure of $P^{-1}P'$ and have
been tabulated~\cite{brouwer:96a}.

Labelling the numerator in
Eq.~(\ref{eqNeffMittel}) as $Z$,
\begin{equation}
	Z=\sum_n \betrag{a'_n}^4=\sum_{n,i_1,\ldots,i_4}
		U^\dagger_{n,i_1} U_{n,i_2} U^\dagger_{n,i_3} U_{n,i_4}
		a^*_{i_1}a_{i_2}a^*_{i_3}a_{i_4}\;,
\end{equation}	
and counting
all the permutations of the indices in it 
yields
\begin{equation}
	\langle Z \rangle = 2 (V_{1,1}+V_{2}) \sum_n \sum_{i_1,i_3} \betrag{a_{i_1}}^2 \betrag{a_{i_3}}^2
	\;.
	\label{eqUmittel2}
\end{equation}
For the square,
\begin{equation}
	Z^2= \sum_{n,m} \betrag{a'_n}^4 \betrag{a'_m}^4\;,
\end{equation}
one finds
\begin{multline}
\langle Z^2\rangle = ( 4 V_{1,1,1,1}+24 V_{2,1,1}+32 V_{3,1}+12 V_{2,2}
	+24 V_4 ) \times\\
	\sum_{n\ne m}\sum_{i1,\ldots,i_4} \betrag{ a_{i_1} 
	a_{i_2}a_{i_3}a_{i_4}}^2 \\
	+ ( 24 V_{1,1,1,1}+144 V_{2,1,1}
	+192 V_{3,1}+72 V_{2,2}	+144 V_4 ) \times\\
	\sum_n\sum_{i1,\ldots,i_4} \betrag{ a_{i_1} 
	a_{i_2}a_{i_3}a_{i_4}}^2 \;.
	\label{eqUmittel4}
\end{multline}	

The terms $a_{i_1},\ldots,a_{i_4}$ appearing in the averages above cancel
against the denominator of Eq.~(\ref{eqNeffMittel}). This is as expected since,
as explained, these coefficients depend on the base that was used to determine
them, i.\,e., they depend on the choice of one particular unitary matrix.
Averaging over the unitary group, this choice must be irrelevant, thus the
coefficients drop out, and the results depends only on the number $M$ of modes.

Inserting Eq.~(\ref{eqUmittel2}) into Eq.~(\ref{eqNeffMittel}) and using
the tabulated coefficients from Ref.~\onlinecite{brouwer:96a} gives the
result
\begin{equation}
	\langle \frac{1}{N'_{\mathrm{eff}}}\rangle  = \frac{2}{M+1} \;,
	\label{eqEffMittelwert}
\end{equation}
which has an easy interpretation. The term $M+1$ instead of the naive term $M$
in the denominator is due to the correlations imposed on $U$ as it is unitary,
cf. the factor $M+1$ in Eq.~(\ref{eqCexakt}). The $2$ in the numerator implies
that every mode is occupied with an effective intensity of $1/2$ relative to the
maximum intensity of any mode -- a very plausible result since the intensity is
a random quantity between $0$ and that maximum value.

A similar calculation using Eq.~(\ref{eqUmittel4}) gives for the square
\begin{equation}
\langle \frac{1}{{N'_{\mathrm{eff}}}^2}\rangle  =  4 \frac{M+5}{(M+1)(M+2)(M+3)} \;,
\end{equation}
so that the variance becomes
\begin{equation}
	\var \frac{1}{N'_{\mathrm{eff}}} =  4
	\frac{M-1}{(M+1)^2(M+2)(M+3)}\;.
\end{equation}
	
The variance decreases much faster with $M$ than $1/N'_{\mathrm{eff}}$
or $1/{N'_{\mathrm{eff}}}^2$ as the lowest-orders in $1/M$ cancel when computing
the variance. This implies that $1/N'_{\mathrm{eff}}$ is a self-averaging
quantity, and its value computed from the number $M$ and hence from the etendue
of the radiation field gives a good estimation of the actual value of
$1/N'_{\mathrm{eff}}$. 

One should be aware, however, that this is a statistical statement: in ``almost
all cases'' this statement is correct. By use of particular light sources it
still is possible to have a very different value for $1/N'_{\mathrm{eff}}$. This
is also reflected in the calculation presented here. Equation~(\ref{eqNeff1})
gives the actual value of $1/N_{\mathrm{eff}}$ which, of course, depends on the
values of the $a_n$. The average value from Eq.~(\ref{eqEffMittelwert}) no
longer depends on the $a_n$ and can thus deviate.

\section{Nonorthogonal modes}

The modes $\phi_n(\vec{r})$ of the electromagnetic field are mutually orthogonal
and stationary,  cf. Eq.~(\ref{eqGamma2}), and all the time-dependence of the
electromagnetic field is included in the frequency-dependence of  the
annihilation operators. However, frequently it is rather inconvenient to
describe time-dependence in this way. An extreme example are pulsed lasers
where it is more appropriate to describe the mode structure of each pulse
separately. This comes at a price, though, as the modes of one pulse are then
not necessarily orthogonal to the modes of another pulse. This is primarily an
issue for the static speckle computed in Sec.~\ref{secStaticSpeckle} as the
mutual orthogonality of the modes was essential for the derivation presented
there.

The static speckle contrast in Eq.~(\ref{eqCexakt}) has an easy interpretation:
except for a minor correction due to the unitary of the scattering matrix, every
mode creates a mutually uncorrelated speckle pattern in the detector plane. The
effect of a superposition of different modes is easily understood from this, and
for the static part it is irrelevant if all the modes are excited simultaneously
or rather one after the other. In this section, we will show that the speckle
patterns of two nonorthogonal modes (i.\,e., necessarily at two different
times) are correlated. We will, without loss of generality, refer to these two
times as first and second pulse, respectively.

The first pulse consists of modes $\phi_n$ and annihilation
operators $a_n$.
For a fixed arrangement of mode structure of the light source, scattering
between source and detector, and detector position, thus only averaging
over the fluctuations of the radiation field, the mean photocount
is, completely analogue to Eq.~(\ref{eqAfrequenz3}), 
\begin{equation}
\overline{n}= 
 \eta T \sum_n S_{1n} S^{\dagger}_{n1}
        I_n
        \;.
	\label{eqOrtho1}
\end{equation}   
We now consider a second
set $\phi'_n$ of modes of the light source but keep the 
scattering
between source and detector and the detector position fixed.
The transformation from the $\phi_n$ to the $\phi'_n$ is linear
and can thus be described by a matrix $S'$, 
\begin{equation}
	S'_{kl}=\langle \phi_k | \phi'_l \rangle \;,
	\label{eqOrtho3}
\end{equation}
where the notation $\langle\ldots |\ldots \rangle$ denotes the overlap 
integral. The annihilation operators transform as 
\begin{equation}
	a_n=\sum_k S'_{nk} a'_k \;,
\end{equation}
and the mean photocount thus becomes
\begin{equation}
\bar{n}'= 
 \eta T \sum_n (S S')_{1n} (S S')^{\dagger}_{n1}
        I'_n
        \;.
	\label{eqOrtho2}
\end{equation}   
The essential quantity is the cross-correlation between $\overline{n}$ and
$\bar{n}'$ after averaging over all possible scattering configurations, hence
over all $S$, and only the matrix $S'$ may remain in the result. A positive
correlation means that the light patterns caused by the two radiation fields are
similar. When the photocounts of two such fields are subsequently integrated on
the same detector, the static speckle contrast will be higher than naively
expected by the inverse square-root of number-of-modes law.

The cross-correlation between $\overline{n}$ and $\bar{n}'$ is given by 
\begin{equation}
	c:=\langle \overline{n} \bar{n}'\rangle - \langle \overline{n} 
	\rangle \langle \bar{n}'\rangle \;,
\end{equation}	
and the average has to be taken over the matrix $S$ keeping the matrix $S'$
fixed.
Writing out the first term explicitly,
\begin{equation}
	\overline{n} \bar{n}' = 
	     \sum_{nmkl} S_{1n} S^*_{1n} S_{1k} S^*_{1l} S'_{km}
		S'^*_{lm} I_n I'_m \;,
\end{equation}		
the average yields nonzero terms 
only for $k=l$, and the two cases $k=n$ and $k\ne$ have to be distinguished.
The necessary averages have already been given
in Eq.~(\ref{eqMittelwerteS}), and the result is
\begin{multline}
	\langle \overline{n} \bar{n}' \rangle =
		\frac{2}{M(M+1)} 
		\sum_{nm} \betrag{S'_{nm}}^2 I_n I'_m \\
		+ \frac{1}{M(M+1)}\sum_m \sum_{n\ne k}\betrag{S'_{km}}^2
			I_n I'_m \;.
\end{multline}
This can be simplified by noting that $\sum_{n\ne k}\betrag{S'_{km}}^2
=1-\betrag{S'_{nm}}^2$. The cross-correlator then becomes
\begin{equation}
	c=\frac{1}{M(M+1)} \sum_{nm} \left[ \betrag{
		\langle \phi_n | \phi'_m \rangle}^2
		- \frac{1}{M} \right] I_n I'_m \;.
	\label{eqCrossCorrelator}
\end{equation}	
This equation gives the expected result in the two extreme situations. If all
intensities are equal, the cross-correlation becomes zero since, as explained 
earlier in the context of Eq.~(\ref{eqCexakt}), the static speckle itself
vanishes, and thus also the cross-correlation does. If only a single mode $n$ is
excited, the cross-correlation becomes proportional to $\betrag{\langle \phi_n |
\phi'_n \rangle}^2 - 1/M$. The average value of $\betrag{\langle \phi_n |
\phi'_n \rangle}^2$ when assuming a random $S'$ is equal to $1/M$, and the
cross-correlator is again zero. If $\phi_n$ and $\phi_n'$ are identical,
the cross-correlator becomes maximal. The cross-correlator thus correctly 
quantifies if $\phi_n$ and $\phi'_n$ are stronger correlated than in a random 
configuration, and the overlap integral $\betrag{\langle \phi_n | \phi'_m \rangle}^2$
is directly related to the static speckle pattern.

When a photodetector sums over $N$ pulses, the variance $\var n$ of the
integrated photocount follows from the variances $\var n_k$ of the $k$-th pulse,
given by Eq.~(\ref{eqTemp2exakt}), and from the cross-correlator $c_{kl}$, given
by Eq.~(\ref{eqCrossCorrelator}), between the $k$-th and the $l$-th pulse,
\begin{equation}
	\var n = \sum_k \var n_k + \sum_{k\ne l} c_{kl} \;.
	\label{eqVarMehrere}
\end{equation}

Apart from the rather obvious relation~(\ref{eqVarMehrere}), the results from
this section have another application. Frequently, radiation fields are expanded
not in their modes but rather in a set of ``convenient'' functions, such as
Gaussian beams or plane waves. Drawback of this approach is that this set then
usually has to be overcomplete, hence it has more elements than there had
been modes, and not all its elements are mutually orthogonal. For
example, two Gaussian beams have an overlap that decreases exponentially with
separation but is always larger than zero. Two plane waves restricted to a
finite interval have an overlap that decreases exponentially with the angle
between the two waves but is always finite.

Depending on the way that this expansion is done, some coherence information
might be lost, namely when doing the expansion such that only the time-averaged
intensities at all field-points are matched, and each $\phi_n$ is then assigned
an intensity $I_n$.  This might sound like a bad approximation but effectively
amounts to using the assumptions from Sec.~\ref{secStaticKorrelation}. Inserting
the $I_n$ into Eq.~(\ref{eqTemp2exakt}) would yield a low static speckle
contrast since the number of $I_n$'s is large. However, the correct equation to
use is Eq.~(\ref{eqVarMehrere}) where every term is evaluated for a single
function $\phi_n$,  and the nonorthogonality of the $\phi_n$ implies that the
cross-correlation terms are large. This is as expected since introducing additional
functions $\phi_n$ to describe the same system should not change its computed
properties.

\section{Discussion}

The purpose of this paper has been to give a consistent and concise description
of  multi-frequency and multi-mode effects on photodetection on a quantum level
trying to avoid 
the use of uncontrolled approximations. Its focus lies on exactness, not
on ease of presentation or application to some particular device. It thus
complements previous work, for example by Joseph W.
Goodman or the group around Christer Rydberg~\cite{rydberg:06a,goodman:07}.

In Sec.~\ref{secZeitunschaerfe} we derived how the uncertainty of the
photocount depends on the measurement time and the temporal coherence function.
Among others, it was shown that a description via the coherence time of the
radiation is insufficient to arrive at an exact result. Similarly, in
Sec.~\ref{secStaticSpeckle} not only previously-known approximations of the 
static speckle contrast were retrieved, namely an inverse square-root law with
the
effective number of modes, but in addition also effects of finite etendue
accepted by the ``mixing device'' or of noncoherent, e.\,g. thermal, radiation
followed from the chosen mathematical formalism without additional effort.  

Some of the results of this paper can be explained, albeit only on a qualitative
level, by the ``brick wall'' model: The radiation field is like a brick wall
(horizontally: time, vertically: position), and only bricks in the same column
or row can cause interference effects. Counting the fraction of such bricks
gives an intuitive explanation of the scaling of dynamic and static speckle
contrast upon changing system parameters. If this is sufficient, the methods and
results presented in this paper are a bit of an overkill. If, on the other hand,
exact agreement between predication and performance of some device is essential,
this paper offers the advantage of giving results using only well-defined input
parameters, thus not relying on effective parameters that need to be tuned until
the desired result is retrieved. 

Main function of any theory is to predict whether some device or experimental
setup will work as expected and\,/\,or required. For state-of-the-art
technological applications, the performance might improve by about $25\,\%$ from
one generation of a device or setup to the next -- but rarely more. To predict
if something actually is an improvement, the error due to approximations thus
has to be much smaller this number. As an example, Fig.~\ref{figKurvenformen}
gives an indication of the error made when using the coherence time instead of
the exact shape of the frequency spectrum. If improvement of dynamic speckle is
a topic, one should thus use the exact formulae and refrain from 
approximations.

Depending on the desired application, either the present paper or one of the
previous works by other authors will thus be more appropriate. Furthermore,  we
restrict us to universal problems. For example, we study in
Sec.~\ref{secStaticSpeckle} only the variance of the photocount, hence the
contrast of the dark and bright ``spots'' in the plane of the detector, but not
their spatial correlations, i.\,e., the size of these spots. Reason is that,
while the contrast is universal, the
size of the spots is not and depends on the setup. The book of
Goodman~\cite{goodman:07} is to a large extend devoted to studying  this question
(on the classical level) for a large number of setups of technological or
scientific importance.

In Sec.~\ref{secZeitunschaerfe} the uncertainty of the photocount for known
radiation field has been computed, yielding in addition to the well-known shot
noise term two terms depending on the frequency spectrum and on the photon
correlations of the radiation source, respectively. The importance of the frequency spectrum
is obvious, in particular in view of the two examples presented in this paper.
For the importance of the effects of noncoherent radiation, the situation is
less clear. For traditional thermal light sources, the necessary
information can be found in
this paper; for traditional lasers, the emitted radiation is coherent, and the
question becomes trivial. For modern light sources, such as excimer
lasers~\cite{basting:05} that are increasingly used for high-power applications,
insufficient data is available. Technological advances are fast, and no good
characterisation of the photon correlations for modern devices seems to  have
been published. Since this contribution to the noise inevitably becomes the
dominant term once the measurement time is sufficiently long, this lack of
knowledge is an actual issue. 

The static speckle, computed in Sec.~\ref{secStaticSpeckle}, depends on the
number of excited modes of the radiation field (and the energy distribution
among them) but, in addition, also on the number of modes that could in principle be excited.
This is closely related to the concept of etendue, cf. Sec.~\ref{secEtendue}. 
Even though etendue is a very basic quantity, its microscopic definition does not
seem to have been addressed before. The relation presented between number of
basis function and macroscopic etendue has application beyond this paper as any
modelling of a radiation field -- for the purpose of simulation or for an
analytic study -- needs to start with a given number of modes, and knowing this
number a priori makes this process much more efficient.

While the applicability of the relation between etendue and maximum number of
modes is obvious, this is less the case for Sec.~\ref{secStaticKorrelation}
discussing the relation between etendue and average effective number of modes.
We demonstrated that this average is identical to the actual number for almost
all radiation fields. However, that is ``only'' a  statistical statement: it
assumes that all possible radiation fields with given etendue are equally
likely, and it is still allowed that a few cases (albeit of measure zero)
deviate strongly. Different light sources have different properties, and for any
technological application, one will choose the most appropriate light source.
Selecting a single-mode laser that operates in a high-order Hermite-Gauss mode
would give the most extreme disagreement between the actual number of modes
(=one) and the prediction from Sec.~\ref{secStaticKorrelation}. The results from
this section can thus only be applied when the choice for a particular light
source does not adversely impair the freedom of the  mode structure of the
generated light field. 

\acknowledgments

The author would like to acknowledge valuable discussions with Olaf Dittmann,
Norbert Kerwien and Johannes Wangler.

%\bibliographystyle{apsrev}
%\bibliography{fourier}
%\bibliography{paper}

\end{document}